\documentclass[Times,4pt,aps,pra,twocolumn,showpacs,amsmath,amssymb,floatfix,footinbib,superscriptaddress]{revtex4-2}
\usepackage{amsmath,natbib,graphicx,subfigure,amssymb,graphics,amsmath,mathrsfs,CJK,color,comment}
\usepackage{multirow,fancyhdr,color,bm,tabularx,psfrag,geometry,dcolumn,datetime,physics,booktabs}
\usepackage[colorlinks=true,linkcolor=blue,urlcolor=blue,citecolor=blue]{hyperref}
\usepackage[mathlines]{lineno}
\def\footnoterule{\kern -1mm \hrule width 5.8cm \kern 2.2mm}
\usepackage{tikz,xcolor,hyperref}
\definecolor{lime}{HTML}{A6CE39}
\DeclareRobustCommand{\orcidicon}{%
    \begin{tikzpicture}
    \draw[lime, fill=lime] (0,0)
    circle [radius=0.16]
    node[white] {{\fontfamily{qag}\selectfont \tiny ID}};\draw[white, fill=white] (-0.0625,0.095)
    circle [radius=0.007];
    \end{tikzpicture}
    \hspace{-2mm}}
\foreach \x in {A, ..., Z}
{\expandafter\xdef\csname orcid\x\endcsname{\noexpand\href{https://orcid.org/\csname orcidauthor\x\endcsname}{\noexpand\orcidicon}}}

\begin{document}
\title{Negative refraction with absorption suppressed by electromagneticly induced transparency in a left-handed atomic system}
\author{Shun-Cai Zhao\orcidA{}}
\email[Corresponding author: ]{zhaosc@kust.edu.cn.}
\affiliation{Physics department, Kunming University of Science and Technology, Kunming, 650093, China}

\begin{abstract}
This paper intends to realize negative refraction with absorption suppressed by the electromagneticly induced transparency(EIT) in a dense four-level atomic system. Without the
two equal transition frequencies responding to the probe field, the atomic system displays a negative refraction with the simultaneously negative permittivity and negative permeability(Left-handedness).
The response of the probe field is amplified and propagates transparency in some frequency extents. Therefore, our aim for searching the low-loss negative refraction can be achieved in the scheme, given the main applied limitation of the negative refractive materials is the large amount of dissipation and absorption. However, an excessive signal field intensity would increase the absorption near the resonance in our scheme.
\begin{description}
\item[PACs]{42.50.Gy}
\item[Keywords]{Negative refraction, absorption suppressed, electromagnetic induced transparency(EIT), left-handed}
\end{description}
\end{abstract}
\maketitle

\section{Introduction}

Negative refraction of light, first predicted to occur in materials with simultaneous negative permittivity and permeability in 1968[1],
has attracted considerable attention in the last decade.Materials with negative refraction index promise many surprising and even
counterintuitive electromagnetical and optical effects, such as the reversals of both Doppler shift and Cherenkov effect,negative
refraction[1], amplification of evanescent waves and subwavelength focusing [2-4], negative Goos-H$\ddot{a}$nchen shift[5], quenching
spontaneous emission[6,7]and so on[8]. The negative refraction index materials intrigue the researchers because of their many significant
potential applications. The`` perfect lens ''is one of them. Since a slab of such materials can focus all frequency components of a
two-dimensional image, it may become possible to make a `` perfect lens ''in which imaging resolution is not limited by the diffraction
limit[2]. Up to now, there have been several approaches to the realization of negative refractive index materials, including
artificial composite metamaterials [9,10], photonic crystal structures [11], transmission line simulation[12-13] and chiral
media[14,15] as well as photonic resonant materials(coherent atomic vapour)[16-21]. Despite the wide variety of implementations a major
challenge is the large loss rate of these materials [22-26]. Especially for potential applications such as subdiffraction limit
imaging [2]or electromagnetic cloaking [27-29], the suppression of absorption proofs is crucial[30,31]. Thus, the realization of
negative refraction material without absorption is of great significance. And very recently, some effort[32-35]has been made to realize negative refraction without absorption. K$\ddot{a}$stel et
al[32] realized negative refraction with minimal absorption in a dense atomic gas via electromagnetically induced chirality. The key
ingredient of the scheme is the electromagnetic chirality that
results from coherently coupling a magnetic dipole transition with
an electric dipole transition via atomic coherence induced by the
two-photon resonant Raman transitions. In Ref.[33], K$\ddot{a}$stel
et al also discussed negative refraction with reduced absorption due
to destructive quantum interference in coherently driven atomic
media. The negative refraction with deeply depressing absorption and
without simultaneously requiring both negative electric permittivity
and magnetic permeability (left-handedness)[2]was obtained by F.L.
Li[35]. Ref.[35] shows this at the ideal situation that the two
chirality coefficients have the same amplitude but the opposite
phase.

In this paper we propose a scheme to  realize negative refractive
index with absorption suppressed by the electromagnetic induced
transparency(EIT) and without two equal transition frequencies
responding to the probe field[16,17]. Based on the effect of quantum
coherence, the magnetic response is amplified and the probe field
propagates transparently, the simultaneously negative permittivity
and negative permeability (Left-handedness)can be realized within
the transparency window. Then, the suppression absorption of
left-handed atomic system resulting from the EIT can also be
obtained.

\section{Theory model}

\begin{figure}
\centering
\includegraphics[width=1.2in]{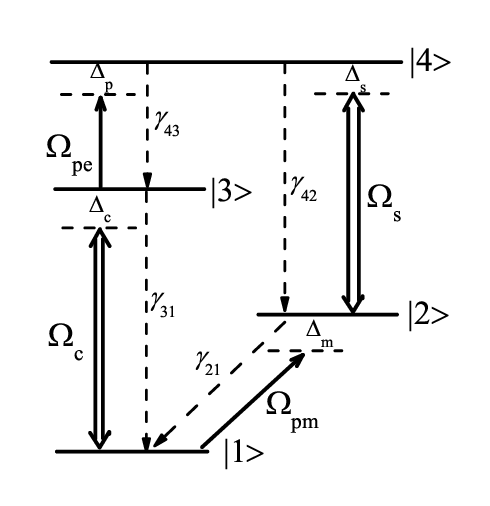}
  \caption{Schematic diagram of a four-level atomic
system interacting with the control $\Omega_{c}$, signal
$\Omega_{s}$ fields and the probe field whose electric and magnetic
components are coupled to the level pairs$|4\rangle-|3\rangle$ and
$|2\rangle-|1\rangle$, respectively.}\label{fig.1}
\end{figure}

The four-level configuration of atoms for consideration is shown in
Figure 1. The parity properties of the atomic states are as follows:
levels $|1\rangle$,$|2\rangle$, and $|3\rangle$ have same parity,
and level $|4\rangle$ is contrary to the three levels. Since the two
lower levels $|1\rangle$ and $|2\rangle$ have same parity and so
$\langle2|$$\hat{\vec{\mu}}$$|1\rangle$$\neq0$ where
$\hat{\vec{\mu}}$ is the magnetic-dipole operator. The two upper
levels,$|3\rangle$ and $|4\rangle$ have the opposite parity with
$\langle4|$$\hat{\vec{d}}$$|3\rangle$$\neq0$ where $\hat{\vec{d}}$
is the electric dipole operator.As shown in Fig.1, three
electromagnetic fields are introduced to couple the four states: The
electric(\textbf{E})and magnetic(\textbf{B}) components of the probe
light(corresponding Rabi frequency
$\Omega_{pe}$$=\frac{\vec{E_{P}}\cdot\vec{d_{34}}}{\hbar}$
,$\Omega_{pm}$$=\frac{\vec{B_{P}}\cdot\vec{\mu_{12}}}{\hbar}$)interact
with the transitions $|3\rangle$ and$|4\rangle$ as well
as$|2\rangle$ and$|1\rangle$, respectively. Hence, the electric and
magnetic components of the probe field with the same frequency
$\omega_{p}$ drive the two transitions $|3\rangle$ -$|4\rangle$ and
$|2\rangle$-$|1\rangle$,simultaneously. The control field with Rabi
frequency denoted by $\Omega_{c}$ pumps atoms in level $|1\rangle$
into upper level $|3\rangle$. According to parity selection rules,
the transition $|1\rangle$-$|3\rangle$ is assumed to be a two-photon
process as stated in [16]The strong signal field couples states
$|2\rangle$ and $|4\rangle$ with Rabi frequency $\Omega_{s}$. In the
rotating-wave and dipole approximations, the Hamiltonian of the
system can be read in the form
\begin{eqnarray*}
H=\sum_{i=1}^{4}\hbar\omega_{i}|i\rangle\langle
i|-\hbar(\Omega_{pm}e^{-i(\omega_{p}t+\theta_{pm})}|2\rangle\langle1|
\\+\Omega_{c}e^{-i(2\omega_{c}t+\theta_{c})}|3\rangle\langle1|
+\Omega_{pe}e^{-i(\omega_{p}t+\theta_{pe})}|4\rangle\langle3|
\\+\Omega_{s}e^{-i(\omega_{s}t+\theta_{s})}|4\rangle\langle2|+c.c.)
\end{eqnarray*}
where $\omega_{i}(i=c,s)$ are the frequencies of the control and
signal fields,respectively. And $\theta_{i}(i=pe,pm,c,s)$ represent
phases of the electric and magnetic components of the probe field,
control and the signal fields, respectively.When the probe field is
weak, i.e.$\Omega_{pe}$,$\Omega_{pm}$
$\ll$$\Omega_{c}$$<$$\Omega_{s}$,we find the first-order
perturbation solution to the Liouville equation in the steady-state
\begin{eqnarray*}
\\\rho_{43}=\frac{1}{{D_{0}D_{1}+D_{2}\Omega_{s}^{2}+\Omega_{s}^{4}}}\{{A_{0}\Omega_{c}^{2}\Omega_{pe}(A_{11}A_{12}+A_{13})}
\\+{e^{i\theta}A_{0}\Omega_{pm}\Omega_{c}\Omega_{s}[A_{21}}
\\-{(\Gamma_{2}+i\Delta_{c})(A_{22}-A_{23}\Omega_{c}^{2}-\gamma_{31}\Omega_{s}^{2})]}\}
\end{eqnarray*}
\begin{eqnarray*}
\\\rho_{21}=\frac{1}{D_{0}D_{1}+D_{2}\Omega_{s}^{2}+\Omega_{s}^{4}}\{{e^{i\theta}A_{0}\Omega_{pe}\Omega_{c}\Omega_{s}\{A_{41}(\Gamma_{2}}
\\+{i\Delta_{c})\Omega_{c}^{2}+(i\Delta_{c}-\Gamma_{2})[\gamma_{31}\Omega_{s}^{2}-A_{42}\Omega_{c}^{2}+(i\Gamma_{6}+\Delta_{c}}
\\-{\Delta_{p})A_{43}]}+{A_{0}\Omega_{pm}A_{31}(i\Delta_{c}-\Gamma_{2})\{A_{33}[(\Gamma_{5}+i\Delta_{p})A_{32}}
\\+{\Omega_{c}^{2}]+A_{32}\Omega_{s}^{2}\}}-{A_{0}\Omega_{pm}\Omega_{c}^{2}(\Gamma_{2}+i\Delta_{c})\{(\gamma_{31}-A_{33})[(\Gamma_{5}}
\\+{i\Delta_{p})A_{32}+\Omega_{c}^{2}]-(A_{32}+\gamma_{31})\Omega_{s}^{2}\}}\}
\end{eqnarray*}
where
\begin{eqnarray*}
\\A_{0}=\frac{i}{\Gamma_{2}^{2}\gamma_{31}+\gamma_{31}\Delta_{c}^{2}+4\Gamma_{2}\Omega_{c}^{2}},\nonumber
\\A_{11}=\gamma_{31}(\Gamma_{2}-i\Delta_{c})+2\Gamma_{1}[\Gamma_{3}+i(\Delta_{c}+\Delta_{p})],\nonumber
\\A_{12}=(\Gamma_{1}+i\Delta_{m})[\Gamma_{6}-i(\Delta_{c}-\Delta_{p})]+\Omega_{c}^{2},\nonumber
\\A_{13}=\Omega_{s}^{2}[i\gamma_{31}\Delta_{c}-\Gamma_{2}(\gamma_{31}-2\Gamma_{6}+2i\Delta_{c}-2i\Delta_{p})],\nonumber
\\A_{21}=(\Gamma_{2}-i\Delta_{c})(\Gamma_{3}+\Gamma_{6}+2i\Delta_{p})(\Gamma_{2}\gamma_{31}+i\gamma_{31}\Delta_{c}+\Omega_{c}^{2}),\nonumber
\\A_{22}=\gamma_{31}(\Gamma_{1}+i\Delta_{m})[-\Gamma_{3}-i(\Delta_{c}+\Delta_{p})],\nonumber
\\A_{23}=\Gamma_{3}-\gamma_{31}+\Gamma_{6}+2i\Delta_{p},A_{31}=-\Gamma_{2}\gamma_{31}-i\gamma_{31}\Delta_{c}-\Omega_{c}^{2},\nonumber
\\A_{32}=\Gamma_{3}+i(\Delta_{c}+\Delta_{p}),A_{33}=\Gamma_{6}+i(\Delta_{p}-\Delta_{c}),\nonumber
\\A_{41}=\Gamma_{3}+\Gamma_{6}+2i\Delta_{p},\nonumber
\\A_{42}=\Gamma_{3}+\gamma_{31}+i(\Delta_{c}+\Delta_{p}),A_{43}=\gamma_{31}(\Delta_{p}-i\Gamma_{5})+i\Omega_{c}^{2},\nonumber
\\D_{0}=(\Gamma_{1}+i\Delta_{m})[\Gamma_{6}-i(\Delta_{c}-\Delta_{p})]+\Omega_{c}^{2},\nonumber
\\D_{1}=(\Gamma_{5}+i\Delta_{p})[\Gamma_{3}+i(\Delta_{c}+\Delta_{p})]+\Omega_{c}^{2},\nonumber
\\D_{2}=(i\Gamma_{6}+\Delta_{c}-\Delta_{p})(\Delta_{p}-i\Gamma_{5})\nonumber
\\+(\Gamma_{1}+i\Delta_{m})[\Gamma_{3}+i(\Delta_{c}+\Delta_{p})]-2\Omega_{c}^{2}.\nonumber
\end{eqnarray*}
with the coherence damping coefficients being given by
$\Gamma_{1}=\frac{1}{2}(\gamma_{1}+\gamma_{21})+\gamma_{c}$,$\Gamma_{2}=\frac{1}{2}(\gamma_{1}+\gamma_{31})+\gamma_{c}$,
$\Gamma_{3}=\frac{1}{2}(\gamma_{1}+\gamma_{42}+\gamma_{43})+\gamma_{c}$,$\Gamma_{4}=\frac{1}{2}(\gamma_{21}+\gamma_{42}+\gamma_{43})+\gamma_{c}$,
$\Gamma_{5}=\frac{1}{2}(\gamma_{31}+\gamma_{42}+\gamma_{43})+\gamma_{c}$,$\Gamma_{6}=\frac{1}{2}(\gamma_{31}+\gamma_{21})$,
in which $\gamma_{c}$ denotes the collisional dephasing rate,and
$\gamma_{1}=0$.The detunings of the applied fields are respectively
defined by $\Delta_{p}=\omega_{p}-(\omega_{4}-\omega_{3})$,
$\Delta_{c}=2\omega_{c}-(\omega_{3}-\omega_{1})$,$\Delta_{s}=\omega_{s}-(\omega_{4}-\omega_{2})$,
$\Delta_{m}=\omega_{p}-(\omega_{2}-\omega_{1})$ and here we set
$\Delta_{m}\neq\Delta_{p}$ (i.e. two equal transition frequencies
aren't required in the atomic system) because
$\Delta_{m}=\Delta_{p}$ is mentioned as a major obstacle in an
actual experimental setting mentioned in Ref.[17]. The relative
phase of the signal, control and probe electric fields and probe
magnetic field is
$\theta=\theta_{pe}+\theta_{c}-\theta_{pm}-\theta_{s}$. It is well
known that the phases of the electric and magnetic components of an
electromagnetic field are identical in a nonconductor [36]. Hence,
in our scheme,$\theta_{pe}=\theta_{pm}$ and the relative phase
becomes the phase difference of the control and pump fields,
i.e.$\theta=\theta_{c}-\theta_{s}$.

In the following, we will discuss the electric and magnetic
responses of the medium to the probe field. When discussing how the
detailed properties of the atomic transitions between the levels are
related to the electric and magnetic susceptibilities, one must make
a distinction between macroscopic fields and the microscopic local
fields acting upon the atoms in the vapor. In a dilute vapor,there
is little difference between the macroscopic fields and the local
fields that act on any atoms(molecules or group of molecules)[36].
But in dense media with closely packed atoms (molecules), the
polarization of neighboring atoms(molecules) gives rise to an
internal field at any given atom in addition to the average
macroscopic field, so that the total fields at the atom are
different from the macroscopic fields[37]. In order to achieve the
negative permittivity and permeability,here the chosen vapor with
atomic concentration$N=5\times10^{24}m^{-3}$ should be dense, so
that one should consider the local field effect, which results from
the dipole-dipole interaction between neighboring atoms. In what
follows we first obtain the atomic electric and magnetic
polarizabilities, and then consider the local field correction to
the electric and magnetic susceptibilities(and hence to the
permittivity and permeability)of the coherent vapor medium. With the
formula of the atomic electric polarizations
$\gamma_{e}=2d_{34}\rho_{43}/\epsilon_{0}E_{p}$,where$E_{p}=\hbar\Omega_{pe}/d_{34}$
one can arrive at
\begin{eqnarray}
\gamma_{e}=\frac{2d_{34}^2\rho_{43}}{\epsilon_{0}\hbar\Omega_{pe}}\
\end{eqnarray}
In the similar fashion, by using the formulae of the atomic magnetic
polarizations $\gamma_{m}=2\mu_{0}\mu_{12}\rho_{21}/B_{p}$ [36], and
the relation between the microscopic local electric and magnetic
fields $E_{p}/B_{p}=c$ we can obtain the explicit expression for the
atomic magnetic polarizability{\color{red}, w}here $\mu_{0}$is the
permeability of vacuum, and c is the speed of light in
vacuum.Then,we have obtained the microscopic physical quantities
$\gamma_{e}$ and $\gamma_{m}$. Thus,{\color{red}$\wedge$}the
coherence $ \rho_{43}$ drives an electric dipole, while the
coherence $\rho_{21}$ drives a magnetic dipole. However, we are
interested in the macroscopic physical quantities such as the
electric and magnetic susceptibilities which are the electric
permittivity and magnetic permeability. The electric and magnetic
Clausius-Mossotti relations can reveal the connection between the
macroscopic and microscopic quantities. According to the
Clausius-Mossotti relation [36], one can obtain the electric
susceptibility of the atomic vapor medium
\begin{eqnarray}
\chi_{e}=N\gamma_{e}\cdot{{{{(1-\frac{N\gamma_{e}}{3})}}}}^{-1}
\end{eqnarray}
The relative electric permittivity of the atomic medium reads
$\varepsilon_{r}=1+\chi_{e}$. The magnetic Clausius-Mossotti [37]
\begin{eqnarray}
\gamma_{m}=\frac{1}{N}(\frac{\mu_{r}-1}{\frac{2}{3}+\frac{\mu_{r}}{3}})
\end{eqnarray}
shows the connection between the macroscopic magnetic permeability
$\mu_{r}$ and the microscopic magnetic polarizations
$\gamma_{m}$.{\color{red}$\wedge$}It follows that the relative
magnetic permeability of the atomic vapor medium is
\begin{eqnarray}
\mu_{r}=\frac{1+\frac{2}{3}N\gamma_{m}}{1-\frac{1}{3}N\gamma_{m}}
\end{eqnarray}
In the above, we obtained the expressions for the electric
permittivity and magnetic permeability of the four-level coherent
atomic vapor system.In the section that follows, we will demonstrate
that under the appropriate parameters condition the permittivity and
permeability of the coherent atomic vapor system can be
simultaneously negative within the transparent window, and the
absorption suppressed deeply results from the EIT.

\section{Results and discussion}

Before doing these calculations,several typical parameters should be
selected. The parameters for the electric and magnetic
polarizabilities of atoms can be chosen as: electric and magnetic
transition dipole moments $d_{21}$=2.5$\times$ $10^{-29}$C$\cdot$m
and $\mu_{34}$=7.0 $\times$$10^{-23}$C$\cdot m^{2}s^{-1}$[17],
respectively.In the model configuration, the transition
from$|2\rangle$ to $|1\rangle$is magnetic dipole allowed and others
are electrical dipole allowed. The spontaneous emission rate of
atomic magnetic dipole transitions is in general smaller than that
of atomic electric dipole transitions by four magnitudes.Thus, in
our numerical calculations, the spontaneous emission rates are
scaled by
$\gamma=10^{7}s^{-1}$:$\gamma_{21}=\gamma_{43}\times(\frac{1}{137})^{2}$[17],
$\gamma_{43}$=0.8$\gamma$, $\gamma_{42}$=1.5$\gamma$,
$\gamma_{31}$=1.2$\gamma$, $\gamma_{c}$=0.8$\gamma$ .In the present
calculations, we choose the density of atoms N to be
$5\times10^{24}m^{-3}$.Dense vapor is required so that the atomic
density should be larger than $10^{24}m^{-3}$[17]. The Rabi
frequency of the probe field is $\Omega_{pe}=0.05\gamma$,and
$\Omega_{c}=8\gamma$ for the control field.The detuning of the
strong signal field is set as $\Delta_{s}=0$ ,and
$\Delta_{c}=\Delta_{m}=0.005\gamma$.The phase difference of the
control and pump fields $\theta$=$\frac{1}{6}\pi$. In Ref.[17], the
probe field hypothesized resonance with the electric and magnetic
transitions is considered a major obstacle in realizing the
predicted effects at an actual experimental setting,as a system with
two states fitting the condition is not directly found . In our
scheme the electric and magnetic transitions are not required to be
resonant with each other.In experimental investigations, the level
configuration shown in Fig. 1 may be realized in the atomic hydrogen
or neon because each has the same level structure as that in Fig.1
[16].

\begin{figure}
  \centering
  \includegraphics[width=0.4\textwidth]{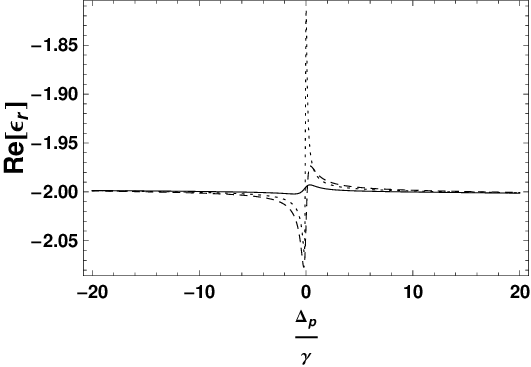} \includegraphics[width=0.4\textwidth]{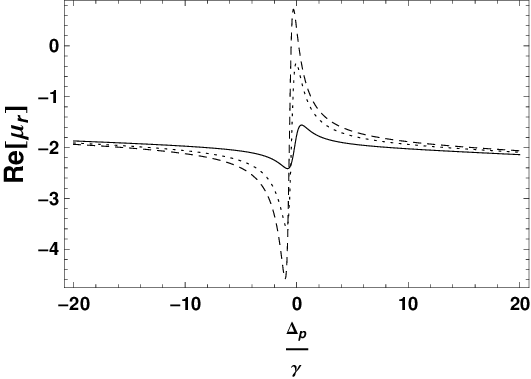}
  \caption{Real parts of the permittivity and
permeability as a function of the rescaled detuning parameter
$\Delta_{p}/\gamma$ and $\Omega_{s}$=14$\gamma$, 18$\gamma$,
20$\gamma$ correspond to the solid, dotted and dashed curves,
respectively. The other parameters are given in the text.}\label{fig.2}
\end{figure}

In the following, we explore the property of both electric
permittivity and magnetic permeability being negative through the
numerical calculations. In Fig.2, the real parts of the relative
electric permittivity $\epsilon_{r}$ and magnetic permeability
$\mu_{r}$ are drawn to find the frequency bands negative
simultaneously for the different values of $\Omega_{s}$=14$\gamma$,
18$\gamma$, 20$\gamma$. Thus the dense atomic system exhibits
left-handedness with simultaneously negative permittivity and
permeability at some frequency extents of the probe field.From the
profiles we observe their values being proximately equal -2 except
near the resonant region. Both the electric and magnetic components
of probe field are amplified, and $\Omega_{s}$=20$\gamma$
corresponds to the most intense magnetic response near the resonant
region.

\begin{figure}
  \centering
  \includegraphics[width=0.42\textwidth]{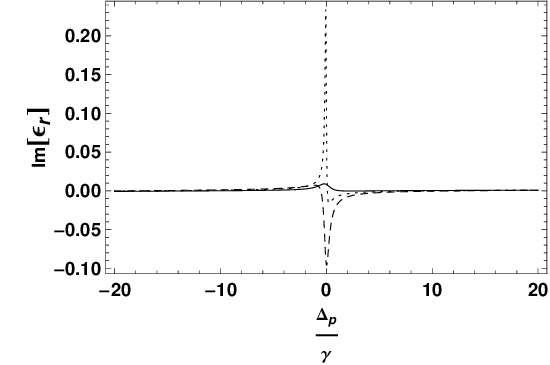}\includegraphics[width=0.42\textwidth]{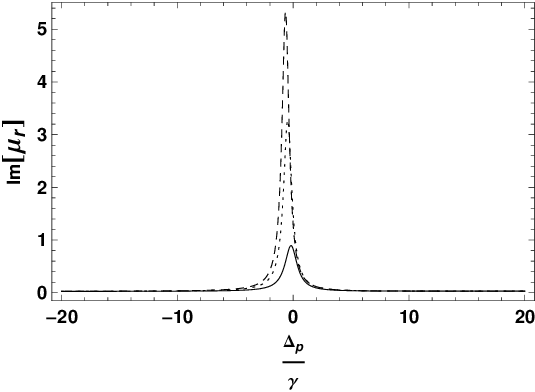}
  \caption{Imaginary parts of the permittivity and
permeability as a function of the rescaled detuning parameter
$\Delta_{p}/\gamma$,and $\Omega_{s}$ =14$\gamma$, 18$\gamma$,
20$\gamma$ correspond to the solid, dotted and dashed curves,
respectively. The other parameters are the same as in Fig. 2.}\label{fig.3}
\end{figure}

The imaginary parts of both the relative electric permittivity
$\epsilon_{r}$ and the relative magnetic permeability $\mu_{r}$ are
plotted in  Fig.3. The electromagnetically induced transparency(EIT)
regions are shown in  Fig.3. This is a very significant result for
us. As is well known, the photon absorption of atom can be greatly
depressed via EIT[38]. And the zero absorption phenomena may occur
in the EIT extents. With observation of the profile of the imaginary
part of $\epsilon_{r}$, the excessive signal field intensity causes
the increasing absorption to gain near the resonant region. However,
the imaginary part of $\mu_{r}$ displays increasing absorption, and
its amplitude is gradually amplified when the signal field varied
the Rabi frequencies by 14$\gamma$, 18$\gamma$ and 20$\gamma$.
Comparing the images of the relative electric permittivity
$\epsilon_{r}$ with the relative magnetic permeability $\mu_{r}$ in
Fig.3, we notice that the magnetic response is stronger than the
electric response of the probe field.

\begin{figure}
  \centering
  \includegraphics[width=0.4\textwidth]{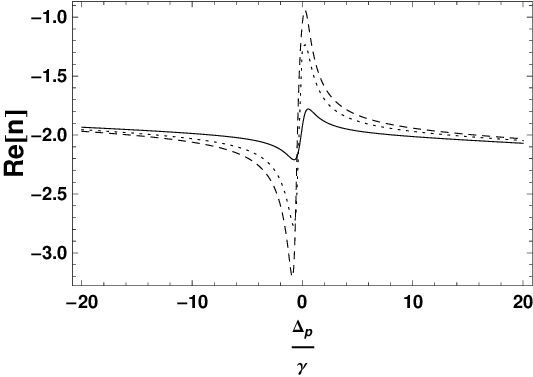} \includegraphics[width=0.4\textwidth]{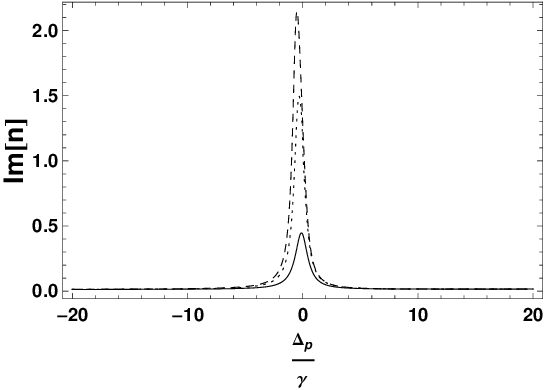}
  \caption{Real and imaginary parts of the refractive
index as a function of the rescaled detuning parameter
$\Delta_{p}/\gamma$,and $\Omega_{s}$=14$\gamma$, 18$\gamma$,
20$\gamma$ correspond to the solid, dotted and dashed curves,
respectively. The other parameters are the same as in Fig.2.}\label{fig.4}
\end{figure}

In Fig.4, the refraction index according to the definition of the
lefthanded material ($n(\omega)=-\sqrt{\epsilon_{r}(\omega)
\mu_{r}(\omega)}$ )[1] is plotted for different values of parameter
$\Omega_{s}$. As observed in Fig.4, the real part of the refractive
index shows negative values and their amplitudes grow gradually with
the increase of $\Omega_{s}$ near the resonant region. The imaginary
part of the refractive index displays absorption enhancing near the
resonant point and absorption depression on both sides, when the
coherent field varied the Rabi frequencies by 14$\gamma$, 18$\gamma$
and 20$\gamma$. The figure of
merit(FOM)($|Re(n)/Im(n)|$)[22,33]shows how much the absorption is
suppressed. When the FOM is much larger than 100, it means that
there is almost no absorption in this area. As shown in Fig.5, the
FOM is far less than 100 in the area near the resonant point. This
illustrates that the increasing absorption occurs in this area,
which is consistent with the peaks of Im[n] near the resonance in
Fig.4. And the FOM has the largest value at the identical re-scaled
detuning parameter $\Delta_{p}/\gamma$ when $\Omega_{s}$=
14$\gamma$. This means that the absorption is depressed deeply when
the signal field varies its Rabi frequency to $\Omega_{s}$=
14$\gamma$, and the excessive signal field intensity wouldn't help
to realize zero absorption. The marked feature of the present scheme
is the absorption depression almost to zero in the left-handed
atomic system, which can understood by the reason for this from the
results shown in Fig.3. Because of the EIT effect, the electric and
magnetic non-absorption extents appear in the re-scaled detuning
parameter $\Delta_{p}/\gamma$ extent, which brings about the zero
value of the imaginary part of refraction index. As a result, the
negative refraction without absorption occurs in the left-handed
atomic system .

\begin{figure}
  \centering
  \includegraphics[width=0.4\textwidth]{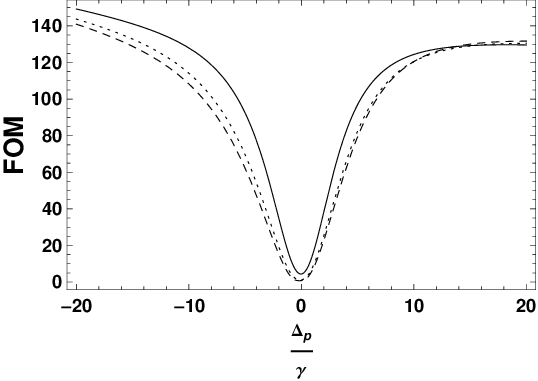}
  \caption{The figure of merit (FOM:$|real(n)/imag(n)|$) as a function
of the detuning parameter detuning $\Delta_{p}/\gamma$,and
$\Omega_{s}$ =14$\gamma$, 18$\gamma$, 20$\gamma$ correspond to the
solid, dotted and dashed curves, respectively. The other parameters
are the same as in Fig. 2.}\label{fig.5}
\end{figure}

\section{Conclusion}

In conclusion, we have demonstrated a scheme for realizing negative
refraction with absorption suppressed by the EIT in the dense
four-level atomic system. Without the two equal transition
frequencies responding to the probe field, the atomic system
displays a negative refraction with left-handedness and vanishing
absorption. The response of the probe field is amplified and
propagates transparently in some frequency extents. Therefore, our
aim for searching the low-loss negative refraction is feasible by
choosing appropriate parameters in the scheme, given that the main
applied limitation of the negative refractive materials is the large
amount of dissipation and absorption. However, an excessive signal
field intensity would increase the absorption near the resonance.
This can also be easily observed here.

\section{Acknowledgment}
 The work is supported by the National
Natural Science Foundation of China (Grant No.60768001 and
No.10464002).

\end{document}